\documentclass[twocolumn,english,prl,showpacs]{revtex4-1}
\usepackage[T1]{fontenc}
\usepackage[latin9]{inputenc}
\usepackage{verbatim}
\usepackage{amsmath}
\usepackage{amssymb}
\usepackage{graphicx}
\usepackage{babel}

\begin{document}

\title{Non-existence of invariant EPR correlation for two qu$\boldsymbol{d}$its,
$\boldsymbol{d}>2$}

\author{Shengshi Pang$^{1}$, Shengjun Wu$^{1}$ and Marek \.{Z}ukowski$^{1,2}$}

\affiliation{$^{1}$ Hefei National Laboratory for Physical Sciences at Microscale
and Department of Modern Physics, University of Science and Technology
of China, Hefei, Anhui 230026, China\\
 $^{2}$ Institute for Theoretical Physics and Astrophysics, University
of Gdansk, PL-80952, Gdansk, Poland}
\begin{abstract}
The two qubit (spin $1/2$) singlet state has invariant perfect EPR
correlations: if two observers measure {\em any} identical observables
the results are always perfectly (anti-)correlated. We show that no
such correlations exists in $d\otimes d$ pure or mixed bipartite
quantum states if $d\geq3$. This points that two qubit singlets are
high-quality quantum information carriers in long distance transmission:
their correlations are unaffected by identical random unitary transformations
of both qubits. Pairs of entangled qu$d$its, $d>2$, do not have
this property. Thus,  qu$d$it properties should be rather exploited
only in protocols  in which transmission is not a long distance one.
\end{abstract}


\maketitle
%\section{Introduction and summary}

The power of entanglement comes from the correlations. Properties
of correlations of a maximally entangled state are purely quantum,
without any classical analog. Recent research dedicated to the nature
of entanglement and quantum correlations is in diverse directions:
e.g., the relation between quantum correlation and classical correlation
\cite{correlation0,correlation1,correlation2,correlation3,correlation4,correlation-1},
quantification of quantum correlations \cite{discord1,correlation4,discord2,discord3,discord4,discord5,discord6},
quantum communication complexity \cite{complexity,quantsim4}.

We shall analyze invariance properties of quantum correlations. A
two qubit singlet has the following property. Imagine two observes,
Alice and Bob, who make measurements on their qubits. Irrespective
of the actual measurement setting they both use, provided they measure
two {\em identical} observables, the results are always perfectly
(anti-)correlated. We shall call such a property {\em invariant
perfect correlation}. It has been used in the Ekert 1991 entanglement
based quantum cryptography protocol \cite{EKERT}, and is important
in many quantum communication schemes. The Bloch Sphere representation
of qubit observables allows one to put them as $\vec{\gamma}^{X}\cdot\vec{\sigma}^{X}$,
where $\vec{\gamma}^{X}$ is a unit real three dimensional vector,
and $\vec{\sigma}^{X}$ is a vector composed of Pauli spin operators,
and $X=A,B$ denotes the observers. If Alice and Bob use identical
measurement settings for their devices we have $\vec{\gamma}^{A}=\vec{\gamma}^{B}$,
and their results fulfill $r_{A}=-r_{B}$, where $r_{X}=\pm1$.

Do we have a similar property for higher dimensional systems, including
higher dimensional singlets? The general answer, which we give here,
is no. One might be surprised that this negative result holds also
for singlets, as results of measurements on higher dimensional singlets
are invariantly correlated in the case of spin observables. However,
such observables form only a subset of all possible ones. Correlations
for other observables, e.g. \cite{ZZH}, behave differently .

This has important ramifications. There is a tendency to investigate
properties of entangled qu$d$its in the hope of getting better realizations
of quantum communication protocols \cite{GISIN}. Such protocols usually
comprise of transfer of the (entangled) objects, manipulations and
finally measurements. Upon a long distance transfer entangled systems
face disturbances. One of them is decoherence, which we shall not
discuss here. However, even if decoherence is avoided, systems may
suffer from random unitary transformations, due to imperfections of
the transmission. The most known are random polarization transformations
of polarization qubits in optical fibers. One of the remedies is to
pass both entangled qubits in a singlet state (almost) simultaneously
via the same transmission line \cite{charlie},
as it was suggested and experimentally
tested by Banaszek et al. \cite{BANASZEK}. The singlet is invariant
with respect to of $U\otimes U$ transformations. This is equivalent
to the property of possessing an invariant perfect correlation discussed
above. Thus its correlations are intact to such a disturbance of the
transmission lines. Is there a similar solution  possible for qu$d$its
($d>2$)? Our results point to a negative answer. There is no two
qu$d$it entangled state, pure or mixed which, has EPR-type correlations
invariant with respect to random, but identical, unitary transformations
applied to both subsystems. This suggests that the qu$d$it properties
should be rather exploited only in quantum information protocols in
which transmission is not a long distance one. { The invariant correlations
of two qubit singlets, single them out to be best for applications
requiring long distance transfer of quantum information. }

In order to substantiate the above claims, let us start with an analysis
of {invariant perfect correlations in two qubit states}.
If
the measurement basis is \emph{arbitrary} but identical for both subsystems,  the measurement results on the two subsystems are \emph{deterministically} EPR correlated and  the correlation type is invariant with
respect to identical changes of the measurement basis on both sides
of the experiment. With respect to spin measurements, represented by Pauli operators,
intuitively, there may exist two types of invariant correlations
in two-qubit states: {type I: $\pm1\leftrightarrow\pm1$,
and {type II:} $\pm1\leftrightarrow\mp1$
where $\leftrightarrow$
denotes the correspondence between the measurement outputs of the
two subsystems.  The following theorem is known to hold for pure states,  we present a version for
arbitrary  states.

\emph{Theorem 1.} \emph{In a $2\otimes2$ composite Hilbert space,
the correlation of type I exists only in the singlet state } $|\psi^{-}\rangle=\frac{1}{\sqrt{2}}(|0^{A}\rangle|1^{B}\rangle-|1^{A}\rangle_{A}|0^{B}\rangle),$
\emph{ and the correlation of type II does not exist in any pure or
mixed state.}

\emph{Proof.} The density operator of an arbitrary $2\otimes2$ bipartite
state can be generally written as
\begin{equation}
\begin{aligned}\rho^{AB}= & \frac{1}{4}[I^{A}\otimes I^{B}+(\boldsymbol{\alpha}\cdot\boldsymbol{\sigma}^{A})\otimes I^{B}+I^{A}\otimes(\boldsymbol{\beta}\cdot\boldsymbol{\sigma}^{B})\\
 & +\sum_{i,j=1}^{3}T_{ij}\sigma_{i}^{A}\otimes\sigma_{j}^{B}].
\end{aligned}
\label{eq:3}
\end{equation}
 The vectors $\boldsymbol{\alpha}$, $\boldsymbol{\beta}$ and components
$T_{ij}$, $i=1,\,2,\,3$, which form the correlation tensor $T$,
are real. One has $||\boldsymbol{\alpha}||\leq1,\,||\boldsymbol{\beta}||\leq1,$
$|T_{ij}|\leq1$, and $||\boldsymbol{\alpha}||^{2}+||\boldsymbol{\beta}||^{2}+||T||^{2}\leq3$

If for all $\vec{n}$ the state $\rho_{AB}$ satisfies $\langle\boldsymbol{n}\cdot\boldsymbol{\sigma}^{A}\otimes\boldsymbol{n}\cdot\boldsymbol{\sigma}^{B}\rangle=-1,$
then the correlation of type I, and if the right hand side is $1$,
the type is II.

By straightforward calculation, we can get the following property
of the correlation function
$
\langle\boldsymbol{n}\cdot\boldsymbol{\sigma}^{A}\otimes\boldsymbol{n}\cdot\boldsymbol{\sigma}^{B}\rangle=\boldsymbol{n}{T}\boldsymbol{n}^{\mathrm{T}},
$
 where ${T}$ is a $3\times3$ matrix with the components of the correlation
tensor $T_{ij}$ as its $ij$-th elements. The symbol $(\cdot)^{\mathrm{T}}$
denotes the transposition of a column vector.

If $\langle\boldsymbol{n}\cdot\boldsymbol{\sigma}^{A}\otimes\boldsymbol{n}\cdot\boldsymbol{\sigma}^{B}\rangle=\pm1$
and the same sign applies to all $\boldsymbol{n}$, one can easily
show that that
$
{T}=\pm{I}_{3}+{X},
$
 where ${I}_{3}$ is the identity matrix of order $3$ and $X$ is
a real anti-symmetric matrix. Thus
\begin{equation}
\begin{aligned}\rho^{AB}= & \frac{1}{4}[I^{A}\otimes I^{B}\pm\sum_{i=1}^{3}\sigma_{i}^{A}\otimes\sigma_{i}^{B}\\
 & +\sum_{1\leq i<j\leq3}X_{ij}(\sigma_{i}^{A}\otimes\sigma_{j}^{B}-\sigma_{j}^{A}\otimes\sigma_{i}^{B})\\
 & +(\boldsymbol{\alpha}\cdot\boldsymbol{\sigma}^{A})\otimes I^{B}+I^{A}\otimes(\boldsymbol{\beta}\cdot\boldsymbol{\sigma}^{B})].
\end{aligned}
\label{eq:5}
\end{equation}

Suppose the eigenvalues of $\rho_{AB}$ in Eq. \eqref{eq:5} are $\lambda_{1}$,
$\lambda_{2}$, $\lambda_{3}$, $\lambda_{4}$, then these four eigenvalues
are the roots of the equation
$
\mathrm{Det}(\lambda I^{A}\otimes I^{B}-\rho^{AB})=0.\
$
 By Vieta's Theorem \cite{Vieta},
we have
\begin{equation}
\sum_{1\leq i<j\leq4}\lambda_{i}\lambda_{j}=-\frac{1}{8}(||\boldsymbol{\alpha}||^{2}+||\boldsymbol{\beta}||^{2}+||X||^{2})\leq0.\label{eq:7}
\end{equation}
 for both {}``$\pm$'' in Eq. \eqref{eq:5}, where $||X||^{2}=\mathrm{Tr}(X^{\dagger}X)$.

According to $\rho^{AB}\geq0$, i.e., $0\leq\lambda_{1},\,\lambda_{2},\,\lambda_{3},\,\lambda_{4}\leq1$,
it can be inferred from \eqref{eq:7} that $\boldsymbol{\alpha}=\boldsymbol{\beta}=\boldsymbol{0}$,
$X=0$, so $\rho^{AB}$ is simplified to
\begin{equation}
\rho^{AB}=\frac{1}{4}(I^{A}\otimes I^{B}\pm\sum_{i=1}^{3}\sigma_{i}^{A}\otimes\sigma_{i}^{B}).\label{eq:8}
\end{equation}

When $\langle\boldsymbol{n}\cdot\boldsymbol{\sigma}^{A}\otimes\boldsymbol{n}\cdot\boldsymbol{\sigma}^{B}\rangle=+1$
for all $\boldsymbol{n}\in\mathbb{R}^{3}$, Eq. \eqref{eq:8} should
take {}``$+$'', but it can be shown that in this case the eigenvalues
of $\rho^{AB}$ are $\pm\frac{1}{2}$, each with multiplicity $2$,
so $\rho^{AB}$ is not positive, and thus no bipartite quantum state
satisfies $\langle\boldsymbol{n}\cdot\boldsymbol{\sigma}^{A}\otimes\boldsymbol{n}\cdot\boldsymbol{\sigma}^{B}\rangle=+1$
for all $\boldsymbol{n}\in\mathbb{R}^{3}$.

When $\langle\boldsymbol{n}\cdot\boldsymbol{\sigma}^{A}\otimes\boldsymbol{n}\cdot\boldsymbol{\sigma}^{B}\rangle=-1$
for all $\boldsymbol{n}\in\mathbb{R}^{3}$, Eq. \eqref{eq:8} should
take {}``$-$'', and it can be shown that in this case the eigenvalues
of $\rho^{AB}$ are $1$ and $0$ (with multiplicity $3$), satisfying
$\rho^{AB}\geq0$, and it is a pure quantum state. By simple calculation,
it can be verified that $\rho^{AB}=|\psi^{-}\rangle\langle\psi^{-}|$,
exactly the singlet state. $\blacksquare$

Theorem 1 is independent of the assumed definition of operators used
in the reasoning. One could claim that type II correlations are possible,
as in a way Bob could interpret his observables as $-\vec{\gamma}_{B}\cdot\vec{\sigma}_{B}$.
But this would lead to inconsistency between operator algebras of
A and B, as $(-\vec{x}\cdot\vec{\sigma})(-\vec{y}\cdot\vec{\sigma})\neq-i\vec{z}\cdot\vec{\sigma}$.
Thus the reached result is objective. This is why a state with permuted
basis states of one of the observers, $\frac{1}{\sqrt{2}}(|0^{A}\rangle|0^{B}\rangle-|1^{A}\rangle_{A}|1^{B}\rangle)$,
does not have invariant perfect correlations.

%\section{Perfect correlation in $d\otimes d$ ($d\geq3$) bipartite quantum states}

Do similar correlations like type I or II exists in bipartite quantum
systems of higher dimensions?

We  define a general \emph{invariant perfect  correlation} for two qu$d$it states as follows: if two measurements along the
same arbitrary orthogonal basis $\{|m_{1}\rangle,\cdots,|m_{d}\rangle\}$
are performed on both subsystems, the outputs of the two measurements
are \emph{deterministically} correlated as a \emph{one-to-one map},
and the map is invariant with respect to  changes of the measurement basis. Different one-to-one
maps between the outputs at the two sides provide a classification
of different types of perfect correlations. Note that, this definition requires
that invariant perfect correlation must be independent of  the labeling
(or, ordering) of the measurement basis states.

Directed graphs, e.g. Fig. 1 and 2, classify the different types of
perfect correlations that may exist in two qu$d$it
states. The states of the measurement basis are represented by vertices.
If two measurement results are correlated, the corresponding vertices
are connected by a directed edge. All directed edges start from the
same subsystem (e.g., $A$) and end at the other one. In other words,
each vertex has double {}``identity'': if it is the starting vertex
of an edge, it denotes a measurement result of subsystem $A$ and
otherwise it denotes a measurement result of subsystem $B$. Such
graphical representation will not cause any confusion or ambiguity
because the edges are directed and the projective measurements on
the two subsystems are along the same basis. The advantage of such
a graphical way to characterize different types of perfect correlations
is that a directed graph defined above stays essentially invariant
under different labeling of the vertices,that is, a perfect correlation
is unchanged under a re-ordering of the states of a measurement basis,
so one \emph{possible} perfect correlation corresponds to one such
directed graph.

Fig. 1 shows all possible types of perfect correlations that may exist
in $2\otimes2$ bipartite quantum states, and Fig. 2 shows all possible
types of perfect correlations that may exist in $3\otimes3$ bipartite
quantum states. It is evident that there can be only two structures
in a directed graph representing perfect correlation: loops (i.e.
a directed edge starts from and ends at the same vertex) or circles.
\begin{figure}
\includegraphics{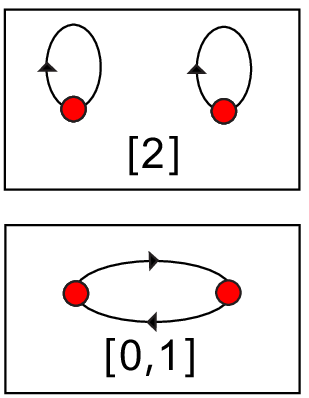}\caption{Possible types of perfect correlations that may exist in $2\otimes2$
bipartite quantum states.}
\end{figure}

\begin{figure}
\includegraphics{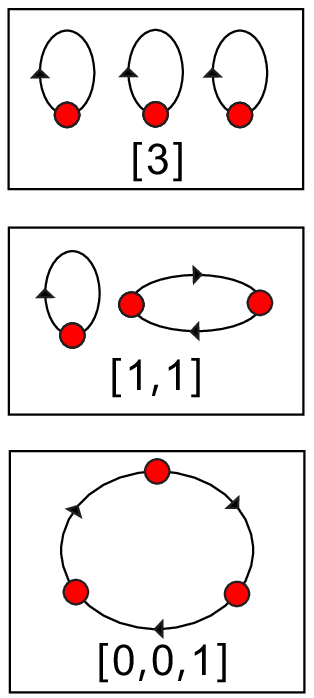}

\caption{Possible types of perfect correlations that may exist in $3\otimes3$
bipartite quantum states}
\end{figure}

We  use a short-hand notation $[n_{1},n_{2},\cdots,n_{w}]$ to denote the
graph with $n_{1}$ loops, $n_{2}$ circles of two vertices, $\cdots$,
$n_{w}$ circles of $w$ vertices, and $w$ is the number of vertices
in the largest circle of the graph. For instance, $[0,1]$ and $[2]$
denote the correlations in type I and type II respectively; and $[d]$
denotes the perfect correlation that measurements on both subsystems
along the same basis always yield the same outcome.

Basing on the above definition of perfect correlation, we set out
to study whether perfect correlation exists in higher dimensional
pure or mixed bipartite quantum states. We  find that
when $d\geq3$ no $d\otimes d$ bipartite quantum state, whether pure
or mixed, has any type of invariant perfect correlations at all, thus
 the two-qubit singlet $|\psi^{-}\rangle$ is the unique
state possessing an invariant perfect correlation.

We first present a lemma.

\emph{Lemma 1. If $d\geq3$, perfect correlations other than $[d]$
cannot exist for any two qu$d$it quantum states.} %%%%%%%%%%%%%%%%%%%%%%%%%%%%%%%%%%%%%%%%%%%%%%%%%%%%%%%%%%%%%%%%%%%%%%%%%%%%%

\emph{Proof.} Suppose there exists a $d\otimes d$ bipartite state
$\rho^{AB}$ which has perfect correlation other than $[d]$. For
projective measurements, on the two subsystems , along the same orthogonal
basis $\{|e_{1}\rangle,|e_{2}\rangle,\cdots,|e_{d}\rangle\}$ one
has
\begin{equation}
\langle e_{1}^{A}|\rho^{AB}|e_{1}^{A}\rangle=c|e_{p}^{B}\rangle\langle e_{p}^{B}|,\, p\neq1\, c>0.\label{eq:20}
\end{equation}
 However, we can change the basis with respect to which the state
is defined, to another orthogonal basis $\{|e_{1}\rangle,|e_{2}^{\prime}\rangle,\cdots,|e_{d}^{\prime}\rangle\}$
which has only one state $|e_{1}\rangle$ in common with the first
basis $\{|e_{1}\rangle,|e_{2}\rangle,\cdots,|e_{d}\rangle\}$, i.e.,
$|e_{i}\rangle\langle e_{i}|\neq|e_{j}^{\prime}\rangle\langle e_{j}^{\prime}|$,
$\forall i,j=2,\cdots,d$. Then, according to the definition of perfect
correlation,
\begin{equation}
\langle e_{1}^{A}|\rho^{AB}|e_{1}^{A}\rangle=c^{\prime}|e_{q}^{\prime B}\rangle\langle e_{q}^{\prime B}|,\, q\neq1\, c^{\prime}>0,\label{eq:21}
\end{equation}
 but $|e_{p}^{B}\rangle\neq|e_{q}^{\prime B}\rangle$, so Eq. \eqref{eq:21}
contradicts Eq. \eqref{eq:20}. Thus, perfect correlations other than
$[d]$ cannot exist in $d\otimes d$ ($d\geq3$) systems. $\blacksquare$

Note that the type of basis change used in the proof is possible only
for $d>2$. This is why the perfect correlation $[0,1]$ exists for
the $2\otimes2$ state $|\psi^{-}\rangle$.

Note that, Lemma 1 does not guarantee that perfect correlation of
identity maps does exist in some $d\otimes d$ ($d\geq3$) quantum
states. Actually we have the following.

\emph{Theorem 2. When $d\geq3$, no $d\otimes d$ pure or mixed bipartite
states has any type of perfect correlations.}

\emph{Proof for the pure states.} According to Lemma 1, we only need
to consider the perfect correlation $[d]$. Assume that a $d\otimes d$
state $|\Phi\rangle$ has such a correlation. Thus, for an arbitrary
basis $\{|e_{1}\rangle,\cdots,|e_{d}\rangle\},$ $|\Phi\rangle$ should
have the following form:
\begin{equation}
|\Phi\rangle=\alpha_{1}|e_{1}\rangle|e_{1}\rangle+\cdots+\alpha_{d}|e_{d}\rangle|e_{d}\rangle,\;\alpha_{1},\cdots,\alpha_{d}\in\mathbb{C}.\label{eq:18}
\end{equation}
 Let us change the measurement basis to $\{|f_{1}\rangle,|f_{2}\rangle,\cdots,|f_{d}\rangle\}$
such that $\{|f_{1}\rangle,\,|f_{2}\rangle\}$ spans the same subspace
spanned by $\{|e_{1}\rangle,|e_{2}\rangle\}$. The invariant perfect
correlation $[d]$ for $|\Phi\rangle$, implies that if the measurement
in the new basis on one of the subsystems produces an output linked
with$|f_{1}\rangle$ or $|f_{2}\rangle$, the second measurement will
produce an identical output. Thus, the two-dimensional (unnormalized)
state
\begin{equation}
|\phi\rangle=\alpha_{1}|e_{1}\rangle|e_{1}\rangle+\alpha_{2}|e_{2}\rangle|e_{2}\rangle\label{eq:19}
\end{equation}
 has perfect correlation $[2]$, i.e. type II, which is a contradiction
with the property of $|\psi^{-}\rangle$. $\blacksquare$

\emph{Proof for mixed states.} Again, Lemma 1 tells that a mixed bipartite
quantum state $\rho_{AB}$ can have an invariant perfect correlation
only of type $[d]$. For measurements on the two subsystems of $\rho_{AB}$
are along the same basis, e.g. \emph{$\{|e_{1}\rangle,\cdots,|e_{d}\rangle\}$},
we have
\begin{equation}
\mathrm{Prob}(i,j)=\langle e_{i}^{A}|\langle e_{j}^{B}|\rho^{AB}|e_{i}^{A}\rangle|e_{j}^{B}\rangle=0,\quad\forall i\neq j.\label{eq:23}
\end{equation}
 Thus, the eigenstates of $\rho^{AB}$ lie in the subspace spanned
by $\{|e_{i}^{A}\rangle|e_{i}^{B}\rangle$, $i=1,\cdots,d\}$. Suppose
the rank of $\rho_{AB}$ is $r$, and let
\begin{equation}
|g_{k}^{AB}\rangle=\sum_{i}\gamma_{k,i}|e_{i}^{A}\rangle|e_{i}^{B}\rangle,\; k=1,\cdots,r,\label{eq:24}
\end{equation}
 be the eigenstates of $\rho_{AB}$, then
\begin{equation}
\rho^{AB}=\sum_{k=1}^{r}\lambda_{k}|g_{k}^{AB}\rangle\langle g_{k}^{AB}|.\label{eq:25}
\end{equation}
 Thus, $\rho_{AB}$ must have the form
\begin{equation}
\rho_{AB}=\sum_{i,j=1}^{d}\alpha_{i,j}|e_{i}^{A}\rangle|e_{i}^{B}\rangle\langle e_{j}^{A}|\langle e_{j}^{B}|,\label{eq:26}
\end{equation}
 where $\alpha_{i,j}=\sum_{k}\lambda_{k}\gamma_{k,i}\gamma_{k,j}^{*}$.

Let us take an arbitrary orthogonal basis as the measurement basis
on both subsystems. We choose a state of the basis $|m\rangle=\sum_{k=1}^{d}m_{i}|e_{k}\rangle.$
The invariant perfect correlation $[d]$ implies that if the measurement
on the subsystem $A$ produces the output linked with $|m^{A}\rangle$,
the state of the subsystem $B$ collapses to $|m^{B}\rangle$. Therefore,
\begin{equation}
\begin{aligned}\langle m^{A}|\rho^{AB}|m^{A}\rangle & =\sum_{i,j,k,l=1}^{d}m_{k}^{*}m_{l}\alpha_{i,j}\delta_{i,k}\delta_{j,l}|e_{i}^{B}\rangle\langle e_{j}^{B}|\\
 & =\sum_{i,j=1}^{d}m_{i}^{*}m_{j}\alpha_{i,j}|e_{i}^{B}\rangle\langle e_{j}^{B}|=c_{|m\rangle}|m^{B}\rangle\langle m^{B}|\\
 & =c_{|m\rangle}\sum_{i,j=1}^{d}m_{i}^{*}m_{j}|e_{i}^{B}\rangle\langle e_{j}^{B}|,
\end{aligned}
\label{eq:30}
\end{equation}
 where $c_{|m\rangle}$ is the probability that the measurement produces
the output $|m\rangle$. Since Eq. \eqref{eq:30} should hold true
for arbitrary $|m\rangle$, we have $\alpha_{i,j}=c_{|m\rangle}.$
Therefore,
\begin{equation}
\begin{aligned}\rho_{AB} & =c_{|m\rangle}\sum_{i,j=1}^{d}|e_{i}^{A}\rangle|e_{i}^{B}\rangle\langle e_{j}^{A}|\langle e_{j}^{B}|\\
 & =c_{|m\rangle}(\sum_{i=1}^{d}|e_{i}^{A}\rangle|e_{i}^{B}\rangle)(\sum_{j=1}^{d}\langle e_{j}^{A}|\langle e_{j}^{B}|).
\end{aligned}
\label{eq:32}
\end{equation}
 Thus, $\rho_{AB}$ is a pure bipartite quantum state, contradicting
the assumption that $\rho_{AB}$ is a mixed state. $\blacksquare$

Combining Theorem 1 and 2, we finally reach:

\emph{Theorem 3. Among all $d\otimes d$ ($d\geq2$) bipartite quantum
states, pure or mixed, only the two-qubit singlet state has an invariant
perfect correlation (it is of type $[0,1]$).}

Thus our claim is substantiated. Note that, Theorem 3 does not state
that there are no EPR correlations in bipartite states other than
$|\psi^{-}\rangle$. Our results consider only the invariance property
of the correlations. Such studies could be extended to multi-qu$d$it
systems. We conjecture that this would single out various types of
multiqubit singlets \cite{CABELLO} as possessing invariant correlations.
It is worth noting that such singlets are already within experimental
reach, see \cite{MOHAMED}.

Supported by the NNSF of China (Grant 11075148), the Fundamental Research
Funds for the Central Universities, the CAS and the National Fundamental
Research Program. SP also acknowledges the Innovation Foundation of
USTC. MZ acknowledges VII UE FP project Q-ESSENCE, and a MNiSW grant
N202 208538.


\begin{thebibliography}{References}
\bibitem{charlie}For other direct benefits of such a scheme see the following. Suppose Charlie is a quantum resource provider who sells
correlated
pairs to other people. If the correlation is not invariant, then Charlie has to tell the buyers the
directions along which they can get the perfect EPR correlations. But information on the
directions
requires an  infinite number bits to be encoded precisely, and this has to be passed to the buyers. If
the
correlation is invariant, such a problem can be avoided.
\bibitem{Vieta}Vieta's theorem characterizes the relation between
the coefficients and the roots of a polynomial equation as follows:
suppose $x^{n}+a_{1}x^{n-1}\cdots+a_{n-1}x+a_{n}=0$ is a polynomial
equation defined on the complex field $\mathbb{C}$ and it has $n$
roots denoted by $x_{1},\cdots,x_{n}$, then there must be $\sum_{1\leq i_{1}<i_{2}<\cdots<i_{k}\leq n}x_{i_{1}}x_{i_{2}}\cdots x_{i_{k}}=\left(-1\right)^{k}a_{k}$.
\end{thebibliography}
\end{document}